\documentclass[12pt,a4paper]{article}
\usepackage{latexsym}
\begin{document}

\begin{flushright}
CERN--TH/97--161\\
{\bf hep-th/9707113}\\
July $17$th, $1997$
\end{flushright}

\begin{center}

%title

{\large {\bf A Note on the D-2-Brane of the Massive Type~IIA Theory 
and Gauged Sigma Models}}

\vspace{.9cm}

%authors
{\large
{\bf Tom\'as Ort\'{\i}n}
\footnote{E-mail address: {\tt Tomas.Ortin@cern.ch}}${}^{,}$
\footnote{Address after October 1997:  {\it IMAFF, CSIC, Calle de Serrano 
121, E-28006-Madrid, Spain}}\\
\vspace{.4cm}
{\it C.E.R.N.~Theory Division}\\
{\it CH--1211, Gen\`eve 23, Switzerland}\\
}

\vspace{.8cm}

%%%%%%%%%%%%%%%%%%%%%%%%%%%%%%%%%%%%%%%%%%%%%%%%%%%%%%%%%%%%%%%%%%%%%%

{\bf Abstract}

\end{center}

\begin{quotation}

\small

Gauging the M-2-brane effective action with respect to an Abelian
isometry in such a way that the invariance under gauge transformations
of the 3-form potential is maintained (slightly modified) we obtain a
fully covariant action with 11-dimensional target space that gives the
massive D-2-brane effective action upon dimensional reduction in the
direction of the gauged isometry.

\end{quotation}

\vspace{3cm}

\begin{flushleft}
CERN--TH/97--161\\
\end{flushleft}

\newpage

\pagestyle{plain}

%%%%%%%%%%%%%%%%%%%%%%%%%%%%%%%%%%%%%%%%%%%%%%%%%%%%%%%%%%%%%%%%%%%%%%

\section*{Introduction}

Recently a dual effective action for the D-2-brane of the massive
type~IIA theory has been proposed in Ref.~\cite{kn:L}. The worldvolume
theory has an extra scalar instead of the Born-Infeld worldvolume
field. In the massless limit, the scalar corresponds to the eleventh
coordinate of the M-2-brane upon dimensional reduction. The
equivalence between the usual action with Born-Infeld vector field and
the dual action was established via an intermediate action that we
present here with an auxiliary worldvolume metric for convenience:

\begin{equation}
\label{eq:intermediate}
\begin{array}{rcl}
S_{\rm I}\ [X^{\mu},X,V_{i},C_{i},\gamma_{ij}]= \hspace{-3cm} &  & \\
& & \\
& & 
-\frac{T_{M-2}}{2} \int d^{3}\xi\ \sqrt{|\gamma|} 
\left\{ \gamma^{ij} 
\left[ e^{-\frac{2}{3}\phi}g_{ij} -
e^{\frac{4}{3}\phi} F_{i}F_{j} \right]
-1\right\} \\
& & \\
& & 
+\frac{T_{M-2}}{3!}\int d^{3}\xi\ \epsilon^{ijk}
\left\{C_{ijk} +6\pi \alpha^{\prime}D_{i}X{\cal F}_{jk}
+6m(\pi\alpha^{\prime})^{2}V_{i}\partial_{j}V_{k} \right\}\, .
\end{array}
\end{equation}

Here

\begin{equation}
\left\{
\begin{array}{rcl}
F_{i} & = & D_{i}X +A^{(1)}{}_{i}\, ,\\
& & \\
D_{i}X & = & \partial_{i}X + C_{i}\, ,\\
& & \\
{\cal F}_{ij} & = & 2\partial_{[i}V_{j]} 
-\frac{1}{2\pi \alpha^{\prime}}B^{(1)}{}_{ij}\, .\\
\end{array}
\right.
\end{equation}

This action has two extra worldvolume fields. If we eliminate $C_{i}$
by using its equation of motion

\begin{equation}
F_{i} = -2\pi \alpha^{\prime} e^{-\frac{4}{3}\phi}\ 
{}^{\star}{\cal F}_{i}\, ,  
\end{equation}

\noindent in the intermediate action, one gets

\begin{equation}
\label{eq:usual}
\begin{array}{rcl}
S\ [X^{\mu},V_{i},\gamma_{ij}] =\hspace{-2.5cm} &  &  \\
& & \\
& & 
-\frac{T_{M-2}}{2} \int d^{3}\xi\ \sqrt{|\gamma|} 
\left\{ \gamma^{ij} 
e^{-\frac{2}{3}\phi}g_{ij} 
+2(\pi \alpha^{\prime})^{2} e^{\frac{4}{3}\phi}
\gamma^{ij} \gamma^{kl} {\cal F}_{ik}{\cal F}_{jl}
-1\right\} \\
& & \\
& & 
+\frac{T_{M-2}}{3!}\int d^{3}\xi\ \epsilon^{ijk}
\left\{C_{ijk} -6\pi \alpha^{\prime}A^{(1)}{}_{i}{\cal F}_{jk}
+6m(\pi\alpha^{\prime})^{2}V_{i}\partial_{j}V_{k} \right\}\, .
\end{array}
\end{equation}

This is the usual action for the D-2-brane of the massive type~IIA
theory and here we clearly see that $V_{i}$ is the characteristic
Born-Infeld vector field and ${\cal F}$ is its field strength.  The
scalar $X$ present in the intermediate action has completely
disappeared in favour of its dual $V_{i}$ at the same time we
eliminated $C_{i}$.

If instead we eliminate $V_{i}$ using its equation of motion

\begin{equation}
\partial_{[i}V_{j]} = -{\textstyle\frac{1}{m\pi \alpha^{\prime}}}
\partial_{[i}C_{j]}\, ,
\end{equation}

\noindent we get, up to a total derivative

\begin{equation}
\label{eq:dual}
\begin{array}{rcl}
\tilde{S}\ [X^{\mu},X,C_{i},\gamma_{ij}]= \hspace{-3cm} &  & \\
& & \\
& & 
-\frac{T_{M-2}}{2} \int d^{3}\xi\ \sqrt{|\gamma|} 
\left\{ \gamma^{ij} 
\left[ e^{-\frac{2}{3}\phi}g_{ij} -
e^{\frac{4}{3}\phi} F_{i}F_{j} \right]
-1\right\} \\
& & \\
& & 
+\frac{T_{M-2}}{3!}\int d^{3}\xi\ \epsilon^{ijk}
\left\{C_{ijk} +3D_{i}XB^{(1)}{}_{jk}
-\frac{6}{m}C_{i}\partial_{j}C_{k} \right\}\, .
\end{array}
\end{equation}

In this action, which, as we have shown, is, classically, completely
equivalent to the usual action Eq.~(\ref{eq:usual}) $V_{i}$ has
completely disappeared in favour of its dual, the scalar field $X$, but
an extra, auxiliary, worldvolume field, $C_{i}$ remains.  The main
interest of this action is that it is very close to the action that
one gets by direct dimensional reduction of the M-2-brane
Ref.~\cite{kn:T} and in fact it reduces to it when the massless limit
is appropriately taken as explained in Ref.~\cite{kn:L}. This could be
helpful in finding the 11-dimensional origin of the massive type~IIA
theory. Here we will present an action with 11-dimensional target
space that gives Eq.~(\ref{eq:dual}) upon direct dimensional
reduction.

The fact that $C_{i}$ is auxiliary can be read out from the symmetries
that the above action enjoys. Apart from worldvolume and target-space
reparametrizations we have gauge transformations of the RR potentials

\begin{equation}
\left\{
\begin{array}{rcl}
\delta_{\Lambda^{(1)}}X & = & -\Lambda^{(1)}(X^{\nu})\, ,\\
& & \\
\delta_{\Lambda^{(1)}} A^{(1)}{}_{\mu} & = & 
\partial_{\mu}\Lambda^{(1)}\, ,\\
& & \\
\delta_{\Lambda^{(1)}} C_{\mu\nu\rho} & = & 
+3B^{(1)}{}_{[\mu\nu}\partial_{\rho]}\Lambda^{(1)}\, ,\\
& & \\
\delta_{\chi} C_{\mu\nu\rho} & = & 3\partial_{[\mu}\chi_{\nu\rho]}\, ,\\
\end{array}
\right.
\end{equation}

\noindent local shifts of the scalar $X$\footnote{This is the symmetry 
  gauged by $C_{i}$. Its presence is necessary in order to have the
  right number of bosonic degrees of freedom, namely 8. The scalar $X$
  can be completely eliminated by using the equation of motion of the
  auxiliary vector field $C_{i}$: $F_{i}= \frac{4}{m}
  e^{-\frac{4}{3}\phi}\ {}^{\star}G_{i}$ where $G_{ij}
  =2\partial_{[i}C_{j]} +\frac{m}{2}B^{(1)}{}_{ij}$ is the
  gauge-invariant field strength of $C_{i}$. When $X$ is eliminated
  $C_{i}$ becomes a dynamical field and, in fact, if we substitute
  $C_{i}=-m\pi \alpha^{\prime}V_{i}$ we recover precisely the action
  Eq.~(\ref{eq:usual}). This is yet another proof of the classical
  equivalence between the dual action Eq.~(\ref{eq:dual}) and the
  usual Eq.~(\ref{eq:usual}).}

\begin{equation}
\left\{
\begin{array}{rcl}
\delta_{\eta}X & = & \eta (\xi)\\
& & \\
\delta_{\eta}C_{i} & = & -\partial_{i}\eta\, ,\\
\end{array}
\right.
\end{equation}

\noindent and the ``massive gauge transformations''\footnote{Here the 
invariance is up to a total derivative.}

\begin{equation}
\left\{
\begin{array}{rcl}
\delta_{\lambda}A^{(1)}{}_{\mu} & = & m \lambda_{\mu}\, ,\\
& & \\
\delta_{\lambda}C_{i} & = & -m \lambda_{i}\, ,\\
\end{array}
\right.
\hspace{1.5cm}
\left\{
\begin{array}{rcl}
\delta_{\lambda}B^{(1)}{}_{\mu\nu} & = & 
-4\partial_{[\mu}\lambda_{\nu]}\, ,\\
& & \\
\delta_{\lambda} C_{\mu\nu\rho} & = & 3m 
\lambda_{[\mu}B^{(1)}{}_{\nu\rho]}\, .\\
\end{array}
\right.
\end{equation}

Observe that the action is not invariant under gauge transformations
of the NSNS 2-form $B^{(1)}_{\mu\nu}$. This is, however, expected
because this field appears in the supergravity Lagrangian with a mass
term that breaks the corresponding gauge invariance \cite{kn:R}.

In Ref.~\cite{kn:L} it was suggested that this auxiliary vector field
could be interpreted in a similar fashion as the auxiliary vector
field introduced in Ref.~\cite{kn:BJO2} to write a covariant action
for the 11-dimensional Kaluza-Klein monopole. In the next section we
will propose a different interpretation for the auxiliary $C_{i}$
vector field which is, though, much in the same spirit.

%%%%%%%%%%%%%%%%%%%%%%%%%%%%%%%%%%%%%%%%%%%%%%%%%%%%%%%%%%%%%%%%%%%%%%

\section{Gauging the M-2-Brane Effective Action}
\label{sec-gauging}

The (bosonic part of) the M-2-brane is \cite{kn:BST}

\begin{equation}
\label{eq:M2}
\begin{array}{rcl}
\hat{S}\ [\hat{X}^{\hat{\mu}},\gamma_{ij}] & = & 
-\frac{T_{M-2}}{2}\int d^{3}\xi\ \sqrt{|\gamma|} 
\left\{ \gamma^{ij} \partial_{i}\hat{X}^{\hat{\mu}}
\partial_{j}\hat{X}^{\hat{\nu}} \hat{g}_{\hat{\mu}\hat{\nu}} -1
\right\}\\
& & \\
& & 
+\frac{T_{M-2}}{3!}\int d^{3}\xi\ \epsilon^{ijk}
\partial_{i}\hat{X}^{\hat{\mu}}
\partial_{j}\hat{X}^{\hat{\nu}}
\partial_{k}\hat{X}^{\hat{\rho}}
\hat{C}_{\hat{\mu}\hat{\nu}\hat{\rho}}\, .\\
\end{array}
\end{equation}

This action is invariant under worldvolume reparametrizations,
11-dimensional spacetime reparametrizations\footnote{Here we write
  $\delta_{0}\phi=\phi^{\prime}(x) -\phi (x)\neq
  \phi^{\prime}(x^{\prime}) -\phi (x)=\delta\phi$.}

\begin{equation}
\left\{
\begin{array}{rcl}
\delta_{\hat{\epsilon}}\ \hat{X}^{\hat{\mu}} & = & 
\hat{\epsilon}^{\hat{\mu}}(\hat{X})\, ,\\
& & \\
\delta_{0\ \hat{\epsilon}}\ \hat{g}_{\hat{\mu}\hat{\nu}} & = & 
-\pounds_{\hat{\epsilon}}\ \hat{g}_{\hat{\mu}\hat{\nu}}
+\hat{\epsilon}^{\hat{\lambda}}\partial_{\hat{\lambda}}
\hat{g}_{\hat{\mu}\hat{\nu}}\, , \\
& & \\
\delta_{0\ \hat{\epsilon}}\ \hat{C}_{\hat{\mu}\hat{\nu}\hat{\rho}}
& = & -\pounds_{\hat{\epsilon}}\ \hat{C}_{\hat{\mu}\hat{\nu}\hat{\rho}}
+\hat{\epsilon}^{\hat{\lambda}}\partial_{\hat{\lambda}}
\hat{C}_{\hat{\mu}\hat{\nu}\hat{\rho}}\, , \\
\end{array}
\right.
\end{equation}

\noindent and gauge transformations of the 3-form potential

\begin{equation}
\delta_{\hat{\chi}}\hat{C}_{\hat{\mu}\hat{\nu}\hat{\rho}}
=3\partial_{[\hat{\mu}}\hat{\chi}_{\hat{\nu}\hat{\rho}]}\, .
\end{equation}

Let us now consider the infinitesimal transformations 

\begin{equation}
\left\{
\begin{array}{rcl}
\delta_{\eta}\hat{X}^{\hat{\mu}} & = & 
\eta (\xi)\hat{k}^{\hat{\mu}}(\hat{X})\, ,\\
& & \\
\delta_{\eta}\hat{g}_{\hat{\mu}\hat{\nu}} & = & 
\eta\hat{k}^{\hat{\lambda}}\partial_{\hat{\lambda}}
\hat{g}_{\hat{\mu}\hat{\nu}}\, , \\
& & \\
\delta_{\eta} \hat{C}_{\hat{\mu}\hat{\nu}\hat{\rho}}
& = & 
\eta \hat{k}^{\hat{\lambda}}\partial_{\hat{\lambda}}
\hat{C}_{\hat{\mu}\hat{\nu}\hat{\rho}}\, , \\
\end{array}
\right.
\end{equation}

\noindent where $\eta$ is an infinitesimal worldvolume scalar 
and $\hat{k}^{\hat{\mu}}$ an arbitrary spacetime vector field. The
action is not invariant under them (they are {\it not}
reparametrizations) but transforms according to

\begin{equation}
\begin{array}{rcl}
\delta_{\eta}\hat{S} & = & 
-\frac{T_{M-2}}{2}\int d^{3}\xi\ \sqrt{|\gamma|} 
\gamma^{ij} \left\{ 2\partial_{i}\eta\hat{k}_{i}+
\eta \partial_{i}\hat{X}^{\hat{\mu}}
\partial_{j}\hat{X}^{\hat{\nu}} 
\pounds_{\hat{k}} \hat{g}_{\hat{\mu}\hat{\nu}} \right\}\\
& & \\
& & 
\hspace{-1cm}
+\frac{T_{M-2}}{3!}\int d^{3}\xi\ \epsilon^{ijk}
\left\{ 3\partial_{i}\eta \left(i_{\hat{k}}\hat{C}\right)_{jk}
+\eta
\partial_{i}\hat{X}^{\hat{\mu}}
\partial_{j}\hat{X}^{\hat{\nu}}
\partial_{k}\hat{X}^{\hat{\rho}}
\pounds_{\hat{k}}\hat{C}_{\hat{\mu}\hat{\nu}\hat{\rho}}
\right\}\, ,\\
\end{array}
\end{equation}

\noindent where

\begin{equation}
\left(i_{\hat{k}}\hat{C}\right)_{\hat{\mu}\hat{\nu}}=
\hat{k}^{\hat{\rho}} \hat{C}_{\hat{\mu}\hat{\nu}\hat{\rho}}\, .
\end{equation}

This transformation will be a symmetry if $\eta$ is constant and
$\pounds_{\hat{k}} \hat{g}_{\hat{\mu}\hat{\nu}} = \pounds_{\hat{k}}
\hat{C}_{\hat{\mu}\hat{\nu}\hat{\rho}}=0$. We will assume that the two
latter conditions hold (so, in particular, $\hat{k}^{\hat{\mu}}$ is a
Killing vector and the metric has an isometry) and will try to modify
the action in order to make it invariant under the transformations
with non-constant $\eta$, i.e.~we will gauge the symmetry present for
constant $\eta$. 

The first step is to substitute everywhere in the original action
(\ref{eq:M2}) the partial derivatives of the $\hat{X}^{\hat{\mu}}$
fields by covariant derivatives:

\begin{equation}
D_{i}\hat{X}^{\hat{\mu}} =\partial_{i}\hat{X}^{\hat{\mu}} 
+C_{i}(\xi) \hat{k}^{\hat{\mu}}\, ,
\end{equation}

\noindent where $C_{i}$ is an auxiliary (non-dynamical) worldvolume 
field that transforms as follows

\begin{equation}
\delta_{\eta}C_{i}=-\partial_{i}\eta\, ,
\Rightarrow  \delta_{\eta} D_{i}\hat{X}^{\hat{\mu}}
=\eta D_{i}\hat{X}^{\hat{\nu}}
\partial_{\hat{\nu}}\hat{k}^{\hat{\mu}}\, .
\end{equation}

We get

\begin{equation}
\label{eq:M2gauged0}
\begin{array}{rcl}
\hat{S}_{{\rm gauged}_{0}}\ [\hat{X}^{\hat{\mu}},C_{i},\gamma_{ij}] & = & 
-\frac{T_{M-2}}{2}\int d^{3}\xi\ \sqrt{|\gamma|} 
\left\{ \gamma^{ij} D_{i}\hat{X}^{\hat{\mu}}
D_{j}\hat{X}^{\hat{\nu}} \hat{g}_{\hat{\mu}\hat{\nu}} -1
\right\}\\
& & \\
& & 
+\frac{T_{M-2}}{3!}\int d^{3}\xi\ \epsilon^{ijk}
D_{i}\hat{X}^{\hat{\mu}}
D_{j}\hat{X}^{\hat{\nu}}
D_{k}\hat{X}^{\hat{\rho}}
\hat{C}_{\hat{\mu}\hat{\nu}\hat{\rho}}\, .\\
\end{array}
\end{equation}

While this substitution automatically makes the action invariant under
the $\delta_{\eta}$ transformations, it completely breaks the gauge
invariance of the 3-form potential\footnote{In principle one could
  have made the discussion using the field strength as in
  Ref.~\cite{kn:HuS}.}:

\begin{equation}
\label{eq:variation}
\delta_{\hat{\chi}}  \hat{S}_{{\rm gauged}_{0}} =  
{\textstyle\frac{T_{M-2}}{3!}}\int d^{3}\xi\ \epsilon^{ijk} 
\left\{ 3C_{i}\left(\pounds_{\hat{k}}\hat{\chi}\right)_{jk} 
-12C_{i} \partial_{j}\hat{\lambda}_{k}\right\}\, ,
\end{equation}

\noindent where

\begin{equation}
\label{eq:lambda}
-{\textstyle\frac{1}{2}}\hat{\lambda}_{\hat{\mu}}\equiv
\left(i_{\hat{k}}\hat{\chi}\right)_{\hat{\mu}} 
=\hat{k}^{\hat{\nu}} \hat{\chi}_{\hat{\mu}\hat{\nu}}\, .
\end{equation}

While the first term in $\delta_{\hat{\chi}} \hat{S}_{{\rm gauged}_{0}}$
can be set to zero by restricting the admissible gauge transformation
to those with parameter $\hat{\chi}_{\hat{\mu}\hat{\nu}}$ whose Lie
derivative with respect to $\hat{k}^{\hat{\mu}}$ vanish (which is
understandable as a condition to preserve the previous condition
$\pounds_{\hat{k}} \hat{C}_{\hat{\mu}\hat{\nu}\hat{\rho}}=0$), it is
not immediately clear how to cancel the second term. The solution,
inspired by the dual D-2-brane action Eq.~(\ref{eq:dual}) is to make
the auxiliary field to transform under $\delta_{\hat{\chi}}$:

\begin{equation}
\delta_{\hat{\chi}}C_{i}=-m\hat{\lambda}_{i}\, ,    
\end{equation}

\noindent and to add a Chern-Simons (CS) term (which is automatically 
gauge-invariant under $\delta_{\eta}$) to the Wess-Zumino (WZ) term of
the action $\hat{S}_{\rm gauged\ 0}$, getting

\begin{equation}
\label{eq:M2gauged}
\begin{array}{rcl}
\hat{S}_{\rm gauged}\ [\hat{X}^{\hat{\mu}},C_{i},\gamma_{ij}] & = & 
-\frac{T_{M-2}}{2}\int d^{3}\xi\ \sqrt{|\gamma|} 
\left\{ \gamma^{ij} D_{i}\hat{X}^{\hat{\mu}}
D_{j}\hat{X}^{\hat{\nu}} \hat{g}_{\hat{\mu}\hat{\nu}} -1
\right\}\\
& & \\
& &
\hspace{-1.5cm}
+\frac{T_{M-2}}{3!}\int d^{3}\xi\ \epsilon^{ijk} 
\left\{
D_{i}\hat{X}^{\hat{\mu}}
D_{j}\hat{X}^{\hat{\nu}}
D_{k}\hat{X}^{\hat{\rho}}
\hat{C}_{\hat{\mu}\hat{\nu}\hat{\rho}}
-\frac{6}{m}C_{i}\partial_{j}C_{k}\right\}\, .\\
\end{array}
\end{equation}

The introduction of the parameter $m$ with dimensions of inverse
length is unavoidable to make the CS term dimensional correct. Its
value is, from the classical point of view, immaterial.  On the other
hand, the effect of the introduction of the CS term is (as we will
see) to add one degree of freedom to the action: without the CS term
both $C_{i}$ and one coordinate scalar could be eliminated by using
the equation of motion of $C_{i}$ (one degree of freedom is subtracted
in the gauging). With the CS term $C_{i}$ does not disappear in doing
this and the number of degrees of freedom is identical to that of the
original action.

However, even though we now are able to cancel exactly the variation
Eq.~(\ref{eq:variation}) we have introduced a new variation and we get
new non-vanishing result with terms coming both from the
kinetic\footnote{Observe that the covariant derivative transforms also
  covariantly under this transformation: $\delta_{\hat{\chi}}D_{i}
  \hat{X}^{\hat{\mu}}= -m D_{i} \hat{X}^{\hat{\nu}}
  \hat{\lambda}_{\hat{\nu}}\hat{k}^{\hat{\mu}}$.}  and WZ terms.
Fortunately this can be cancelled without any further modification of
the action by simply modifying the gauge transformation of $\hat{C}$
and making the metric transform under $\delta_{\hat{\chi}}$. Thus, to
summarize, we find that the gauged action Eq.~(\ref{eq:M2gauged}) is
invariant (up to a total derivative) under the $\delta_{\hat{\chi}}$
transformations

\begin{equation}
\left\{
\begin{array}{rcl}
\delta_{\hat{\chi}} \hat{C}_{\hat{\mu}\hat{\nu}\hat{\rho}}
& = & 3\partial_{[\hat{\mu}}\hat{\chi}_{\hat{\nu}\hat{\rho}]}
+3m \hat{\lambda}_{[\hat{\mu}}
\left(i_{\hat{k}}\hat{C} \right)_{\hat{\nu}\hat{\rho}]}\, , \\
& & \\
\delta_{\hat{\chi}} \hat{g}_{\hat{\mu}\hat{\nu}} & = & 
2m \hat{\lambda}_{(\hat{\mu}}\hat{k}_{\hat{\nu})}
=2m \hat{\lambda}_{(\hat{\mu}}
\left(i_{\hat{k}}\hat{g}\right)_{\hat{\nu})}\, ,\\
& & \\
\delta_{\hat{\chi}}C_{i} & = & -m\hat{\lambda}_{i}\, , \\ 
\end{array}
\right.
\end{equation}

\noindent and the $\delta_{\eta}$ transformations

\begin{equation}
\left\{
\begin{array}{rcl}
\delta_{\eta}\hat{X}^{\hat{\mu}} & = & 
\eta (\xi)\hat{k}^{\hat{\mu}}(\hat{X})\, ,\\
& & \\
\delta_{\eta}C_{i} & = & -\partial_{i}\eta\, ,\\
\end{array}
\right.
\hspace{1.5cm}
\left\{
\begin{array}{rcl}
\delta_{\eta}\hat{g}_{\hat{\mu}\hat{\nu}} & = & 
\eta\hat{k}^{\hat{\lambda}}\partial_{\hat{\lambda}}
\hat{g}_{\hat{\mu}\hat{\nu}}\, , \\
& & \\
\delta_{\eta} \hat{C}_{\hat{\mu}\hat{\nu}\hat{\rho}}
& = & 
\eta \hat{k}^{\hat{\lambda}}\partial_{\hat{\lambda}}
\hat{C}_{\hat{\mu}\hat{\nu}\hat{\rho}}\, , \\
\end{array}
\right.
\end{equation}

\noindent assuming the conditions

\begin{equation}
\pounds_{\hat{k}} \hat{g}_{\hat{\mu}\hat{\nu}} 
=\pounds_{\hat{k}} \hat{C}_{\hat{\mu}\hat{\nu}\hat{\rho}}
=\pounds_{\hat{k}} \hat{\chi}_{\hat{\mu}\hat{\nu}}=0\, ,
\end{equation}

\noindent hold.

%%%%%%%%%%%%%%%%%%%%%%%%%%%%%%%%%%%%%%%%%%%%%%%%%%%%%%%%%%%%%%%%%%%%%%

\section{Direct Dimensional Reduction}
\label{sec-reduction}

It is easy to see that the above action Eq.~(\ref{eq:M2gauged}) gives
the dual action Eq.~(\ref{eq:dual}) of the D-2-brane of the massive
type~IIA theory proposed in Ref.~\cite{kn:L} upon direct dimensional
reduction in the direction $X$ associated to the isometry we have
gauged.  Choosing coordinates adapted to the isometry
$\hat{k}^{\hat{\mu}}=\delta^{\hat{\mu}x}$ we split the eleven
coordinates $\hat{X}^{\hat{\mu}}$ into the ten 10-dimensional
$X^{\mu}\, ,\,\,\,\mu=0,\ldots,9$ and the extra scalar
$X=\hat{X}^{10}$.

Using the relations between the 11-dimensional and 10-dimensional
fields

\begin{equation}
\left\{
\begin{array}{rcl}
\hat{g}_{xx} & = & -e^{\frac{4}{3}\phi}\, ,\\
& & \\
\hat{g}_{\mu x} & = & \left(i_{\hat{k}}\hat{g}\right)_{\mu} = 
-e^{\frac{4}{3}\phi}A^{(1)}{}_{\mu}\, , \\
& & \\
\hat{g}_{\mu\nu} & = & e^{-\frac{2}{3}\phi} g_{\mu\nu} 
-e^{\frac{4}{3}\phi}A^{(1)}{}_{\mu}A^{(1)}{}_{\nu}\, , \\
\end{array}
\right.
\hspace{.5cm}
\left\{
\begin{array}{rcl}
\hat{C}_{\mu\nu\rho} & = & C_{\mu\nu\rho}\, , \\
& & \\
\hat{C}_{\mu\nu x} & = & \left(i_{\hat{k}}\hat{C} \right)_{\mu\nu}=
B^{(1)}{}_{\mu\nu}\, ,\\
\end{array}
\right.
\end{equation}

\begin{equation}
\left\{
\begin{array}{rcl}
\hat{\chi}_{\mu\nu} & = & \chi_{\mu\nu}\, , \\
& & \\
\hat{\lambda}_{\mu} & = & \lambda_{\mu}\, ,\\
\end{array}
\right.
\end{equation}

\noindent (observe that $\hat{\lambda_{x}} =\hat{k}^{\hat{\mu}} 
\hat{\lambda}_{\hat{\mu}} =0$) it is immediate to recover both the
dual action and all its symmetries. Observe that the transformations
$\delta_{\chi}$ and $\delta_{\lambda}$ become independent in ten
dimensions.

%%%%%%%%%%%%%%%%%%%%%%%%%%%%%%%%%%%%%%%%%%%%%%%%%%%%%%%%%%%%%%%%%%%%%%

\section{Conclusion}
\label{sec-conclusion}

We have constructed an effective action for a membrane with
11-dimensional target space whose direct dimensional reduction gives
the dual action for D-2-brane of the massive 10-dimensional type~IIA
theory found in Ref.~\cite{kn:L}.

The action has a natural interpretation as a gauged sigma model in
which the gauge invariance of the 3-form potential has been preserved
in a slightly modified way. At the same time, the total number of
scalar degrees of freedom has been preserved, making the
supersymmetrization of the action (assuming $\kappa$-symmetry)
possible.

On the other hand, the present action may help us in understanding the
11-dimensional origin of the massive type~IIA theory. There is no
cosmological 11-dimensional supergravity theory (see
Ref.~\cite{kn:BDHS} and references therein) and our result may
indicate how to accommodate the solitons of the massive 10-dimensional
type~IIA theory into the usual 11-dimensional supergravity theory. In
this sense, gauged sigma models may proof to be a useful tool
\cite{kn:BJO3}.

%%%%%%%%%%%%%%%%%%%%%%%%%%%%%%%%%%%%%%%%%%%%%%%%%%%%%%%%%%%%%%%%%%%%%%

\section*{Acknowledgments}

I would like to acknowledge most interesting correspondence with
E.~Bergshoeff and Y.~Lozano. I am also indebted to M.M.~Fern\'andez
for her support.

%%%%%%%%%%%%%%%%%%%%%%%%%%%%%%%%%%%%%%%%%%%%%%%%%%%%%%%%%%%%%%%%%%%%%%%

\end{document}